\documentstyle[12pt]{article}
\def\beq{\begin{equation}}
\def\eeq{\end{equation}}
\def\bea{\begin{eqnarray}}
\def\eea{\end{eqnarray}}
\def\nn{\nonumber}
\def\ba{\begin{array}}
\def\ea{\end{array}}
\font\twelvemsbm=msbm10 at 12 true pt
\font\eightmsbm=msbm8
\font\sevenmsbm=msbm7
\newfam\msbmfam
\textfont\msbmfam=\twelvemsbm
\scriptfont\msbmfam=\eightmsbm
\scriptscriptfont\msbmfam=\sevenmsbm
\def\Bbb#1{{\fam\msbmfam\relax#1}} 
\def\R{\Bbb R}
\def\C{\Bbb C}
\def\one{1\hskip -1mm{\rm l}}
\begin{document}
\addtolength{\baselineskip}{2mm}
\renewcommand{\thefootnote}{\fnsymbol{footnote}}
\setcounter{footnote}{1}

\begin{center}
{\Large \bf Finite dimensional representations of \\
the quantum group $GL_{p,q}(2)$ using \\
the exponential map from $U_{p,q}(gl(2))$}

\vspace{2cm}

{\bf R. Jagannathan\footnote{Permanent address: The Institute of
Mathematical Sciences, Madras - 600113, India;
E-mail address: jagan@imsc.ernet.in}
and
J. Van der Jeugt\footnote{Senior Research Associate of N.F.W.O
(National Fund for Scientific Research of Belgium);
E-mail address: Joris.VanderJeugt@rug.ac.be}}

\vspace{0.5cm}

Department of Applied Mathematics and Computer Science \\
University of Ghent, Krijgslaan 281-S9, B-9000 Gent, Belgium
\end{center}

\vspace{2cm}

\noindent
{\bf Abstract}:  Using the Fronsdal-Galindo formula for the exponential
mapping from the quantum algebra $U_{p,q}(gl(2))$
to the quantum group $GL_{p,q}(2)$, we show how the $(2j+1)$-dimensional
representations of $GL_{p,q}(2)$ can be obtained by `exponentiating' the
well-known $(2j+1)$-dimensional representations of $U_{p,q}(gl(2))$ for
$j$ $=$ $1,\frac{3}{2},\ldots $;
$j$ $=$ $\frac{1}{2}$ corresponds to the
defining 2-dimensional $T$-matrix.  The earlier results on the
finite-dimensional representations of $GL_q(2)$ and $SL_q(2)$
(or $SU_q(2)$)
are obtained when $p$ $=$ $q$.  Representations of $U_{\bar{q},q}(2)$
$(q$ $\in$ $\C \backslash \R$ and $U_q(2)$
$(q$ $\in$ $\R \backslash \{0\})$ are also considered.  The structure
of the Clebsch-Gordan matrix for $U_{p,q}(gl(2))$ is studied.
The same Clebsch-Gordan coefficients are applicable in the
reduction of the
direct product representations of the quantum group $GL_{p,q}(2)$.

\vspace{1cm}

\noindent
Short title: Representations of $GL_{p,q}(2)$

\vspace{1cm}

\noindent
Phys. Abs. Class. Nos.~: 0210, 0220, 0365

\newpage

\renewcommand{\theequation}{\arabic{section}.{\arabic{equation}}}
\setcounter{equation}{0}
\section{Introduction}

\noindent
The analysis of the bialgebraic duality relationship between
the pair of Hopf algebras ${\cal A}$ $=$ $A\left(GL_{p,q}(2)\right)$ (or
$Fun_{p,q}(GL(2))$) and ${\cal U}$
$=$ $U_{p,q}(gl(2))$ by Fronsdal and Galindo~\cite{FG} provides the first
example of generalization of the exponential relationship, obtained
between the classical Lie groups and algebras, through the construction of
the form of such a mapping from $U_{p,q}(gl(2))$ to the quantum group
$GL_{p,q}(2)$.  Following~\cite{FG}, Bonechi {\em et al.}~\cite{Bo-et} have
studied the forms of the exponential relationship for a few other examples
of quantum groups, namely, the quantized Heisenberg, Euclidean and Galilei
groups.  More recently, Morozov and Vinet~\cite{MV} have obtained a
generalization of the result of~\cite{FG} for any simple quantum
group with a single deformation parameter.
For $GL_q(2)$, Finkelstein~\cite{F} has obtained the representations
using theory of invariants and studied the converse map $GL_q(2)$
$\longrightarrow$ $U_q(gl(2))$ leading to the identification of the
generators of $U_q(gl(2))$ as the infinitesimal generators of $GL_q(2)$:
hence, the representations of $U_q(gl(2))$ are also derived.

Here, we demonstrate explicitly how the Fronsdal-Galindo theory leads
to a straightforward derivation of the finite-dimensional representaions of
$GL_{p,q}(2)$.  The earlier results on the finite-dimensional irrreducible
representaions of $GL_q(2)$~\cite{F} and $SL_q(2)$
(or $SU_q(2)$)~\cite{W,VS,K,GKK,Ma-et,N,KK}
leading to the $q$-analogues of the classical Wigner $d$-functions or
spherical functions, are seen to follow in the special case when
$p$ $=$ $q$.  We consider briefly also the quantum groups
$U_{\bar{q},q}(2)$ $(q$ $\in$
$\C \backslash \R)$
and $U_q(2)$ $(q$ $\in$ $\R \backslash \{0\})$.

The question of $q$-analogues of the Clebsch-Gordan coefficients (CGCs)
arises naturally when one considers the representation theory of quantum
groups.  The CGCs for $U_q(su(2))$ and $SU_q(2)$ have been discussed
extensively (see~\cite{K,GKK,N,KK}).  In the case of
$GL_{p,q}(2)$~\cite{Ku,De-et,S1,T}, the structure of its dual Hopf algebra,
$U_{p,q}(gl(2))$, is known from studies using different approaches~\cite
{SWZ,D1,D2}. It is clear that one can have only $U_{p,q}(u(2))$
nontrivially and any $(p,q)$-deformation of $U(su(2))$ would depend
effectively only on a single parameter ($\sqrt{pq}$).
So, the earlier studies on
a $(p,q)$-deformed $su(2)$ algebra led naturally to the conclusion
that the corresponding deformed CGCs depend only on $\sqrt{pq}$
(see~\cite{Q,MM}).  Here, we shall arrive at the structure of the
Clebsch-Gordan matrix for $U_{p,q}(gl(2))$ based on Reshetikhin's general
theory of quantum algebras with multiple deformation parameters~\cite{R}.
The same Clebsch-Gordan matrices can be used for the reduction
of direct product representations of the quantum group
$GL_{p,q}(2)$.

Before proceeding to consider the representation theory of $GL_{p,q}(2)$,
in particular, let us first recall some elements of the general theory of
representations of quantum groups.  The Hopf algebra $A(G_q)$,
algebra of functions on a quantum group $G_q$, has a nonabelian coordinate
ring generated by the (variable) elements of a $T$-matrix satisfying the
$RTT$-relation
%
\beq
RT_1T_2 = T_2T_1R \qquad
{\rm with}\ \ \ \ T_1 = T \otimes \one  \ \ \
T_2 = \one \otimes T
\label{RTT}
\eeq
where the $R$-matrix is a solution of the Yang-Baxter equation.  The
coproduct maps of the elements of $T$, $\left\{ T_{mk} \right\}$,
specified by
%
\beq
\Delta\left( T_{mk} \right) = \sum_j\,T_{mj} \otimes T_{jk} \qquad
{\rm or}\ \ \ \
\Delta (T) = T \dot{\otimes} T
\label{copro}
\eeq
provide a unique homomorphism of the $\C$-algebra
generated by $\{ T_{mk} \}$ subject to the relations~(\ref{RTT}).
One may choose a basis for $A(G_q)$ with any convenient
parametrization and take the required coproduct maps as induced from
(\ref{copro}).  A $\C$-vector subspace $V$ spanned by
$\{ V_m$$|$$m = $$1,$ $2,$$\ldots ,$$n$$\}$ of $A(G_q)$
is said to carry an $n$-dimensional representation of $G_q$ if it forms
a left subcomodule of $A$ such that
%
\beq
\Delta (V) = {\cal T} \dot{\otimes} V \qquad
{\rm or}\ \ \ \
\Delta\left( V_m \right) = \sum_{k=1}^n\,{\cal T}_{mk} \otimes V_k \quad
\forall\,m = 1,2,\ldots ,n
\label{defrepV}
\eeq
where $\{ {\cal T}_{mk} \in A \}$ are called the elements of
the `representation matrix' ${\cal T}$.  Then, the comodule structure of
$\{ V_m \}$ implies, as in (\ref{copro}),
%
\beq
\Delta ({\cal T}_{mk}) = \sum_{j=1}^n\,{\cal T}_{mj} \otimes
{\cal T}_{jk} \qquad
{\rm i.e.,}\ \ \ \ \Delta ({\cal T})
= {\cal T} \dot{\otimes} {\cal T}\,.
\label{defrepT}
\eeq
For example, the elements of the first (or, any) column of the $T$-matrix
constitute such a $\C$-vector subspace of
$A(G_q)$ carrying the defining representation with ${\cal T}$
$=$ $T$.   Essentially, a representation matrix ${\cal T}$ of $G_q$ is a
generalization of the $T$-matrix, with $\{ {\cal T}_{mk}$ $\in$
$A(G_q) \}$ and the relation (\ref{defrepT})
satisfied.

The paper is organized as follows.  In section 2, we recall briefly the
main properties of the quantum group $GL_{p,q}(2)$ and its dual Hopf algebra
$U_{p,q}(gl(2))$.  In section 3, we describe the Fronsdal-Galindo method for
exponentiating the representations of $U_{p,q}(gl(2))$ to obtain the
representations of $GL_{p,q}(2)$.  Section 4 gives the explicit
representation matrices $\{{\cal T}\}$ for $GL_{p,q}(2)$ and shows how
the earlier results on the representations of $GL_q(2)$, $SL_q(2)$ and
$SU_q(2)$ are obtained in the appropriate limits.  In section 5 we briefly
remark on the quantum group $U_{\bar{q},q}(2)$.  Section 6 presents the
solution to the problem of CGCs for $U_{p,q}(gl(2))$ and
$GL_{p,q}(2)$. Section 7 contains several concluding remarks.

\section{$GL_{p,q}(2)$ and $U_{p,q}(gl(2))$}
\setcounter{equation}{0}

\noindent
For the quantum group $GL_{p,q}(2)$ the defining $T$-matrix, or the defining
representation matrix, is specified by
%
\beq
T = \left(
\ba{cc}
a & b \\
c & d \\
\ea \right)
\label{defT}
\eeq
with the commutation relations
%
\bea
ab & = & qba \qquad cd = qdc \qquad ac = pca \qquad bd = p db \nn \\
bc & = & (p/q)cb \qquad ad-da = (q-p^{-1})bc
\label{comrel}
\eea
following from the $RTT$-relation (\ref{RTT}) corresponding to
%
\beq
R = Q^{\frac{1}{2}}\left(
\ba{cccc}
Q^{-1} & 0 & 0 & 0 \\
0 & \lambda^{-1} & Q^{-1}-Q & 0 \\
0 & 0 & \lambda & 0 \\
0 & 0 & 0 & Q^{-1}
\ea \right)
\label{R}
\eeq
where
%
\beq
Q = \sqrt{pq} \quad \quad \lambda = \sqrt{p/q}\,.
\label{Qlambda}
\eeq
We are concerned here only with the case of generic values of $Q$,
$\lambda$ $\in$ $\C \backslash \{0\}$.
Also, it may be noted that we shall use both
the sets of notations $(Q,\lambda )$ and $(p,q)$ for the deformation
parameters interchangably, according to convenience, with the relationship
(\ref{Qlambda}) always implied.  The coproduct maps (\ref{copro}) are
explicitly given by
%
\beq
\Delta (T) = \left(
\ba{cc}
\Delta (a) & \Delta (b) \\
\Delta (c) & \Delta (d)
\ea \right)
= \left(
\ba{cc}
a \otimes a + b \otimes c & a \otimes b + b \otimes d \\
c \otimes a + d \otimes c & c \otimes b + d \otimes d
\ea \right)\,.
\label{delT}
\eeq
The quantum determinant of $T$ defined by
%
\beq
{\cal D} = ad - qbc = ad - pcb
\label{qdet}
\eeq
satisfies the commutation relations
%
\beq
{\cal D}a = a{\cal D} \qquad {\cal D}b = \lambda^{-2}b{\cal D} \qquad
{\cal D}c = \lambda^2 c{\cal D} \qquad {\cal D}d = d{\cal D}
\label{qdcomrel}
\eeq
and is a group-like element such that
%
\beq
\Delta ({\cal D}) = {\cal D} \otimes {\cal D}\,.
\label{delqdet}
\eeq

The algebra ${\cal U}$ dual to ${\cal A}$ $=$ $A(GL_{p,q}(2))$,
namely $U_{p,q}(gl(2))$, may be presented in a standard form as follows.
The generators $\{ J_0, J_\pm\,,Z \}$ can be taken to satisfy
the algebra
%
\bea
[ J_0,J_\pm ] & = & \pm J_\pm \quad \quad
[ J_+,J_- ] = [ 2J_0 ]_Q \nn \\
{} [ Z ,J_0 ] & = & 0 \quad \quad
[ Z ,J_\pm ] = 0
\label{alg}
\eea
with
%
\beq
[X]_{\sf q} = \frac{{\sf q}^X-{\sf q}^{-X}}{{\sf q}-{\sf q}^{-1}}
\quad \quad \forall\ \ {\sf q},X\,.
\label{q-no}
\eeq
Whereas the algebraic relations (\ref{alg}) do not depend on $\lambda$,
the coproducts do and are given by, up to equivalence,
%
\bea
\Delta ( J_\pm ) & =
& J_\pm \otimes Q^{-J_0}\lambda^{\pm Z} +
Q^{J_0}\lambda^{\mp Z} \otimes J_\pm \nn \\
\Delta ( J_0 ) & =
& J_0 \otimes \one + \one \otimes J_0 \qquad
\Delta (Z) = Z \otimes \one + \one \otimes Z\,.
\label{coalg}
\eea
The associated universal ${\cal R}$-matrix, relating the coproduct $\Delta$
and the opposite coproduct $\Delta '$ $=$ $\sigma\,\Delta$, with
$\sigma\,(u \otimes v)$ $=$ $v \otimes u$,\,\,$\forall$\,$u,v$
$\in {\cal U}$,
through the equation
%
\beq
\Delta '(u) = {\cal R}\Delta (u){\cal R}^{-1} \qquad
\forall \  u \in\  {\cal U}
\label{univR}
\eeq
is
%
\bea
{\cal R} & = & Q^{-2(J_0 \otimes J_0)}
\lambda^{2(Z \otimes J_0 - J_0 \otimes Z)} \nn \\
&    &  \ \ \times \sum_{n=0}^{\infty}\,
\frac{(1-Q^2)^n}{[n]!}\,Q^{-\frac{1}{2}
n(n-1)}(Q^{-J_0}\lambda^Z J_+
\otimes Q^{J_0}\lambda^Z J_-)^n
\label{calR}
\eea
where
%
\beq
[n] = [n]_Q \qquad
[n]! = [n][n-1]\ldots[2][1] \qquad
[0]! = 1\,.
\label{n!}
\eeq
Hereafter, unless otherwise specified, $[n] = [n]_Q$.
The ${\cal R}$-matrix~(\ref{calR}), say ${\cal R}_{Q,\lambda}$, is easily
obtained~\cite{CJ1} following the observation that, if we denote by
$\Delta_{Q,\lambda}$ the coproduct defined by~(\ref{coalg}), then,
%
\beq
\Delta_{Q,\lambda}(u) = F\Delta_{Q,\lambda =1}(u)F^{-1} \qquad
\forall\  u \in {\cal U} \qquad
{\rm with}\ \
F = \lambda^{(J_0 \otimes Z - Z \otimes J_0)}
\label{FdelF}
\eeq
and by Reshetikhin's theory~\cite{R}
%
\beq
{\cal R}_{Q,\lambda} = F^{-1}{\cal R}_{Q,\lambda =1}F^{-1}\,.
\label{FRF}
\eeq

Here, we are not concerned with the other aspects of the Hopf algebraic
structure, namely, the counit and the antipode.

\section{The exponential map from $U_{p,q}(gl(2))$ to $GL_{p,q}(2)$}
\setcounter{equation}{0}

\noindent
Let us now describe the exponential map from $U_{p,q}(gl(2))$ to
$GL_{p,q}(2)$, {\em \`{a} la} Fronsdal and Galindo~\cite{FG}.  To this
end, we have to redefine the generators of $U_{p,q}(gl(2))$ as
%
\beq
\hat{J}_+ = J_+Q^{-(J_0+\frac{1}{2})}
\lambda^{Z-\frac{1}{2}} \quad
\hat{J}_- = Q^{(J_0+\frac{1}{2})}
\lambda^{Z-\frac{1}{2}}J_- \quad
\hat{J}_0 = J_0 \quad
\hat{Z} = Z
\label{newJ}
\eeq
with the algebra
%
\bea
[ \hat{J}_0,\hat{J}_\pm ] & = & \pm \hat{J}_\pm \quad \quad
[ \hat{J}_+,\hat{J}_-] =
\lambda^{2\hat{Z}-1}[2\hat{J}_0] \nn \\
{} [ \hat{Z},\hat{J}_0 ] & = & 0 \quad \quad
[ \hat{Z},\hat{J}_\pm ] = 0
\label{newalg}
\eea
and the induced coproducts
%
\bea
\Delta (\hat{J}_+) & = & \hat{J}_+ \otimes Q^{-2\hat{J}_0}
\lambda^{2\hat{Z}} + \one \otimes \hat{J}_+ \quad
\Delta (\hat{J}_-) = \hat{J}_- \otimes \one +
Q^{2\hat{J}_0}\lambda^{2\hat{Z}} \otimes \hat{J}_- \nn \\
\Delta (\hat{J}_0) & = & \hat{J}_0 \otimes \one +
\one \otimes \hat{J}_0 \quad \quad
\Delta (\hat{Z}) = \hat{Z} \otimes \one +
\one \otimes \hat{Z}\,.
\label{newcoalg}
\eea

Let the element `$a$' of $T$ be taken to be invertible~\cite{FG} and
expressed as
%
\beq
a = {\rm e}^\alpha\,.
\label{a}
\eeq
Since we are dealing only with nonsingular $T$-matrices we shall
represent
%
\beq
{\cal D} = {\rm e}^{-2\phi} = \xi^{-2}\,.
\label{phi-mu}
\eeq
Further, let us take
%
\beq
\beta = a^{-1}b \qquad \gamma = ca^{-1} \qquad
\delta = \alpha + 2\phi\,.
\label{begadel}
\eeq
The variables $\{ a,b,c,d \}$ can be expressed as
%
\beq
a = {\rm e}^\alpha \qquad b = {\rm e}^\alpha \beta \qquad
c = \gamma {\rm e}^\alpha \qquad
d = \gamma {\rm e}^\alpha \beta + {\rm e}^{-\delta}\,.
\label{a=}
\eeq
The set of new variables $\{ \alpha ,\beta ,\gamma ,\delta \}$ are seen
to form a Lie algebra:
%
\bea
[ \alpha , \beta ] & = & (\rho -\theta )\beta \quad \quad
[ \alpha , \gamma ] = (\rho + \theta )\gamma \nn \\
{} [ \delta , \beta ] & = & (\rho + \theta )\beta \quad \quad
[ \delta , \gamma ] = (\rho - \theta )\gamma \nn \\
{} [ \alpha , \delta ] & = & 0 \quad \quad
[ \beta , \gamma ] = 0 \nn \\
  &   &  \qquad {\rm with}\ \ \ \
Q = {\rm e}^\rho \qquad \lambda = {\rm e}^\theta\,.
\label{Lie}
\eea
Then, following Fronsdal and Galindo~\cite{FG}, one can write down a
`universal ${\cal T}$-matrix' for $GL_{p,q}(2)$ as
%
\beq
{\cal T} = {{\cal E}xp}_{Q^{-2}}\{ \gamma \hat{J}_- \}
{\rm exp}\{ \alpha (\hat{J}_0+\hat{Z}) +
\delta (\hat{J}_0-\hat{Z}) \}
{{\cal E}xp}_{Q^2}\{ \beta \hat{J}_+ \}
\label{calT}
\eeq
where
%
\beq
{{\cal E}xp}_{{\sf q}^2}\{ X \} = \sum_{n=0}^{\infty}\,
\left\{ \prod_{k=1}^{n}({\sf q}^{2k}-1) \right\}^{-1}
({\sf q}^2-1)^n X^n = \sum_{n=0}^{\infty}\,
\frac{{\sf q}^{-\frac{1}{2}n(n-1)}}{[n]_{\sf q}!}\,X^n\,.
\label{calE}
\eeq
The universal ${\cal T}$-matrix (\ref{calT}) embodies all the
finite-dimensional representations of $GL_{p,q}(2)$: substituting the
finite-dimensional numerical matrices representing
$\{ \hat{J}_0,$$\hat{J}_{\pm},\hat{Z} \}$ in the
expression (\ref{calT}) for ${\cal T}$, expanding it, and reexpressing
the resulting matrix elements in terms of $\{a,$$b,$$c,$$d\}$ and
$\xi$ $=$ ${\cal D}^{-\frac{1}{2}}$ using the relations
((\ref{qdet}),(\ref{a})-(\ref{Lie})), one obtains all the
finite-dimensional representation matrices $\{{\cal T}\}$ of
$GL_{p,q}(2)$ in the sense of the representation
theory outlined in the introduction.  Though the definition of the matrix
elements of ${\cal T}$ by (\ref{calT}) involves $a^{-1}$, it is found
possible to express them completely in terms of $\{a,$$b,$$c,$$d\}$ and
the group-like $\xi$ (with $\Delta (\xi )$ $=$ $\xi \otimes \xi )$, using
the relations ((\ref{qdet}),(\ref{a})-(\ref{Lie})).

Except for the additional central element $Z$, the algebra (\ref{alg}) is
same as the standard $U_Q(sl(2))$ for which all the finite-dimensional
irreducible representations are known~(see \cite{Ro,RA}): for generic
values of $Q$$\in$$\C \backslash \{0\}$, these are the
straightforward
$Q$-analogues of the $(2j+1)$-dimensional (spin-$j$) representations of
the classical $sl(2)$, with
$j$ $=$ $0,$$\frac{1}{2},$$1,$$\frac{3}{2},$$\ldots\,.$
So, with the extra central element as $Z$ $=$ $z \one $,
the $(2j+1)$-dimensional irreducible representations of
$\{ \hat{J}_0,$$\hat{J}_\pm ,\hat{Z} \}$ obeying (\ref{newalg}),
say $\left\{ \Gamma^{(\mu )} \mid \mu = (j,z) \right\}$,
are readily obtained (see also~\cite{D3} for more details on the dual Hopf
algebras $A(GL_{p,q}(2))$ and $U_{p,q}(gl(2))$ and representation theory of
quantized universal enveloping algebras $\{ U_q(G) \}$ in general).
Let us call the matrix obtained by substituting the representation
$\Gamma^{(\mu )}$ in the
expression (\ref{calT}) for ${\cal T}$ as ${\cal T}^{(\mu )}$.
The matrix elements are labeled as
$\{ {\cal T}^{(\mu )}_{mk}\mid m,k=j,j-1,\ldots ,-(j-1),-j \}$.

The basic theory underlying the Fronsdal-Galindo formalism is as follows.
Let $\{ x^A \}$ be a basis for $A(G_q)$ such that
%
\beq
x^Bx^C = \sum_{A}\,h^{BC}_A x^A \quad \quad
\Delta (x^C) = \sum_{A,B}\,f^C_{AB} x^A \otimes x^B\,.
\label{xx}
\eeq
Let $\{ X_A \}$ be the basis of $U_q(G)$ chosen such that
%
\beq
X_AX_B = \sum_{C}\,f_{AB}^C X_C \quad \quad
\Delta (X_A) = \sum_{B,C}\,h_A^{BC} X_B \otimes X_C\,.
\label{XX}
\eeq
Then, the universal ${\cal T}$-matrix defined by
%
\beq
{\cal T} = \sum_{A}\,x^AX_A
\label{TauxX}
\eeq
satisfies the equation
%
\beq
\Delta ({\cal T}) = \sum_{A}\,\Delta (x^A) X_A
= {\cal T} \dot{\otimes} {\cal T}
\label{DelcalTau}
\eeq
in view of the duality relations ((\ref{xx}),(\ref{XX})).
Equation (\ref{TauxX}) should be interpreted as defining a universal
${\cal T}$-matrix in the sense that ${\cal T}^{(\mu )}$ $=$
$\sum_{A}\,x^AX^{(\mu )}_A$ $\in$ $A(G_q)$, for
$\{ X^{(\mu )}_A \}$, the numerical representations of
$\{ X_A \}$, in any representation $\Gamma^{(\mu )}$.  Thus, the
elements $\{ {\cal T}^{(\mu )}_{mk}\in A(G_q)\}$
form a representation matrix for the quantum group $G_q$ and the
formula (\ref{TauxX}) expresses abstractly the exponential map $U_q(G)$
$\longrightarrow$ $G_q$.  In the present context of $GL_{p,q}(2)$, the
basis elements of $A(GL_{p,q}(2))$ and $U_{p,q}(gl(2))$
are given (see~\cite{FG} for details of derivation), respectively, by
%
\bea
x^A & = & \gamma^{a_1}\alpha^{a_2}\delta^{a_3}\beta^{a_4} \nn \\
X_A & = &
\frac{Q^{\frac{1}{2}a_1(a_1-1)}\hat{J}_-^{a_1}}
{[a_1]!}
\frac{(\hat{J}_0 + \hat{Z})^{a_2}}{a_2!}
\frac{(\hat{J}_0 - \hat{Z})^{a_3}}{a_3!}
\frac{Q^{-\frac{1}{2}a_4(a_4-1)}\hat{J}_+^{a_4}}
{[a_4]!} \nn \\
    &   &   a_1,a_2,a_3,a_4 = 0,1,2,\ldots\,.
\label{defxX}
\eea

As noted by Bonechi {\em et al.}~\cite{Bo-et} the relations
((\ref{xx}),(\ref{XX})) also imply
%
\beq
{\cal T}_1{\cal T}_2 =
\sum_{A}\,x^A\Delta (X_A) \qquad
{\cal T}_2{\cal T}_1 =
\sum_{A}\,x^A \Delta '(X_A)
\label{calTT}
\eeq
where ${\cal T}_1 = {\cal T} \otimes \one$ and
${\cal T}_2 = \one \otimes {\cal T}$.  Hence, one would have, in the
present case,
%
\beq
R^{(\mu \otimes \mu )}{\cal T}_1^{(\mu )}{\cal T}_2^{(\mu )} =
{\cal T}_2^{(\mu )}{\cal T}_1^{(\mu )}R^{(\mu \otimes \mu )}
\label{RcalTT}
\eeq
in accordance with the FRT-formalism of quantum groups~\cite{FRT} where
$R^{(\mu \otimes \mu )}$ is the $R$-matrix corresponding to the direct
product of two representations $(\Gamma^{(\mu )})$
and obtained by substituting the respective representations in the formula
(\ref{calR}) for the universal ${\cal R}$.

Now, we note that the $R$-matrix (\ref{R}) defining the $T$-matrix
((\ref{defT}),(\ref{comrel})) is
$R^{\left(\frac{1}{2},\frac{1}{2}\right) \otimes \left(\frac{1}{2},
\frac{1}{2}\right)}$.
For a generic value of $z$ $\left( \neq \frac{1}{2} \right)$,
$R^{\left(\frac{1}{2},z\right) \otimes \left(\frac{1}{2},z\right)}$
has the same form as (\ref{R}) with $\lambda$ replaced by $\lambda^{2z}$
and thus defines a $T$-matrix whose elements will obey the commutation
relations of the same form as (\ref{comrel}) with $p$ and $q$ replaced,
respectively, by $p'$ $=$ $p^{z+\frac{1}{2}}/q^{z-\frac{1}{2}}$ and
$q'$ $=$ $q^{z+\frac{1}{2}}/p^{z-\frac{1}{2}}$.  In general, it is clear
{}from (\ref{FdelF}) and (\ref{FRF}) that $R^{(j,z)\otimes (j,z)}$ would
have the same form as $R^{(j,\frac{1}{2})\otimes (j,\frac{1}{2})}$ with
$\lambda$ replaced by $\lambda^{2z}$.

\section{Explicit representation matrices
$\{ {\cal T}^{(j,z)} \}$}
\setcounter{equation}{0}

\noindent
First, let us consider explicitly the matrices
${\cal T}^{\left(\frac{1}{2},\frac{1}{2}\right)}$ and
${\cal T}^{\left(1,\frac{1}{2}\right)}$ and, then, generalize the result.
To verify that ${\cal T}^{\left( \frac{1}{2},\frac{1}{2} \right)}$ is
the defining $T$-matrix (\ref{defT}) we have to substitute the matrices
%
\beq
\hat{J}_+ = \left(
\ba{cc}
0 & 1 \\
0 & 0
\ea \right) \quad
\hat{J}_- = \left(
\ba{cc}
0 & 0 \\
1 & 0
\ea \right) \quad
\hat{J}_0 = \frac{1}{2} \left(
\ba{cr}
1 & 0 \\
0 & -1
\ea \right) \quad
\hat{Z} = \frac{1}{2}\left(
\ba{cc}
1 & 0 \\
0 & 1
\ea \right)
\label{2drep}
\eeq
in (\ref{calT}) and expand.  The result is
%
\bea
{\cal T}^{\left( \frac{1}{2},\frac{1}{2} \right)} & = &
{{\cal E}xp}_{Q^{-2}}\left\{ \left(
\ba{cc}
0 & 0 \\
\gamma & 0
\ea \right) \right\}
{\rm exp}\left\{ \left(
\ba{cr}
\alpha & 0 \\
0 & -\delta
\ea \right) \right\}
{{\cal E}xp}_{Q^2}\left\{ \left(
\ba{cc}
0 & \beta \\
0 & 0
\ea \right) \right\} \nn \\
   &   &  \nn \\
   & = & \left(
\ba{cc}
1 & 0 \\
\gamma & 1
\ea \right) \left(
\ba{cc}
{\rm e}^\alpha & 0 \\
0 & {\rm e}^{-\delta}
\ea \right) \left(
\ba{cc}
1 & \beta \\
0 & 1
\ea \right) =
\left(
\ba{cc}
{\rm e}^\alpha & {\rm e}^\alpha \beta \\
\gamma {\rm e}^\alpha & \gamma {\rm e}^\alpha \beta + {\rm e}^{-\delta}
\ea \right) \nn \\
   &   &  \nn \\
   & = & \left(
\ba{cc}
a & b \\
c & ca^{-1}b+a^{-1}{\cal D}
\ea \right) = \left(
\ba{cc}
a & b \\
c & d
\ea \right)
\label{Tau1/2}
\eea
as expected; this is, in fact, the starting point of the Fronsdal-Galindo
formalism.  It may be noted that at the level of the two-dimensional
defining representation ${\cal E}xp\{\ \}$ can be replaced by ordinary
${\rm exp}\{\ \}$ (see Finkelstein~\cite{F} and Akulov
{\em et al.}~\cite{AGG} for the use of such a realization of the
$2 \times 2$ $T$-matrix for $GL_q(2)$ and $SL_q(2))$.  The deformed
structure of the exponential map (\ref{calT}), brought out by the
Fronsdal-Galindo approach, is revealed only in dimensions $>2$.

For  $\Gamma^{\left(1,\frac{1}{2}\right)}$, with $[2]$ $=$
$Q+Q^{-1}$,
%
\bea
\hat{J}_+ & = & [2]^{\frac{1}{2}}\left(
\ba{ccc}
0 & Q^{-\frac{1}{2}} & 0 \\
0 & 0 & Q^{\frac{1}{2}} \\
0 & 0 & 0
\ea \right) \quad \quad
\hat{J}_- = [2]^{\frac{1}{2}}\left(
\ba{ccc}
0 & 0 & 0 \\
Q^{\frac{1}{2}} & 0 & 0 \\
0 & Q^{-\frac{1}{2}} & 0
\ea \right) \nn \\
   &   &    \nn \\
\hat{J}_0 & = & \left(
\ba{ccr}
1 & 0 & 0 \\
0 & 0 & 0 \\
0 & 0 & -1
\ea \right) \quad \quad
\hat{Z} = \frac{1}{2}\left(
\ba{ccc}
1 & 0 & 0 \\
0 & 1 & 0 \\
0 & 0 & 1
\ea \right)\,.
\label{3drep}
\eea
Substituting this representation in (\ref{calT}) we obtain
%
\beq
{\cal T}^{\left( 1,\frac{1}{2} \right)}
= \left(
\ba{ccc}
{\rm e}^{\phi +2\alpha} &
[2]^{\frac{1}{2}}Q^{-\frac{1}{2}}{\rm e}^{\phi +2\alpha}\beta &
Q^{-1}{\rm e}^{\phi + 2\alpha}\beta^2 \\
   &    &   \\
{}[2]^{\frac{1}{2}}Q^{\frac{1}{2}}
\gamma {\rm e}^{\phi +2\alpha} &
[2]\gamma {\rm e}^{\phi + 2\alpha}\beta + {\rm e}^{-\phi} &
{\ba{c}
([2]^{\frac{1}{2}}Q^{-\frac{1}{2}}\gamma
{\rm e}^{\phi + 2\alpha}\beta^2  \\
 +[2]^{\frac{1}{2}}Q^{\frac{1}{2}}{\rm e}^{-\phi}\beta )
\ea} \\
   &    &    \\
Q\gamma^2{\rm e}^{\phi + 2\alpha} &
{\ba{c}
([2]^{\frac{1}{2}}Q^{\frac{1}{2}}\gamma^2
{\rm e}^{\phi + 2\alpha}\beta   \\
 +[2]^{\frac{1}{2}}Q^{-\frac{1}{2}}\gamma {\rm e}^{-\phi})
\ea} &
{\ba{c}
(\gamma^2{\rm e}^{\phi + 2\alpha}\beta^2  \\
+[2]\gamma {\rm e}^{-\phi}\beta \\
 + {\rm e}^{-(3\phi +2\alpha )})
\ea}
\ea \right)
\label{Tau1alpha}
\eeq
Now, using the relations ((\ref{qdet}),(\ref{a})-(\ref{Lie})), we get
%
\beq
{\cal T}^{(1,\frac{1}{2})} = \xi \left(
\ba{ccc}
a^2 &
[2]^{\frac{1}{2}}Q^{-\frac{1}{2}}ab &
\lambda^{-1}b^2 \\
  &   &  \\
{} [2]^{\frac{1}{2}}Q^{-\frac{1}{2}}ac &
ad+Q^{-1}\lambda^{-1}bc &
[2]^{\frac{1}{2}}Q^{-\frac{1}{2}}\lambda^{-1}bd \\
  &   &  \\
\lambda c^2 &
[2]^{\frac{1}{2}}Q^{-\frac{1}{2}}\lambda cd &
d^2
\ea \right)\,.
\label{Tau1abcd}
\eeq
Note that in deriving (\ref{Tau1abcd}) from (\ref{Tau1alpha}) the
relations (\ref{Lie}) are used in the Heisenberg-Weyl form
%
\bea
{\rm e}^\alpha \beta & = & Q\lambda^{-1} \beta {\rm e}^\alpha \quad
{\rm e}^\alpha \gamma = Q\lambda \gamma {\rm e}^\alpha \quad
{\rm e}^\phi \beta = \lambda \beta {\rm e}^\phi \quad
{\rm e}^\phi \gamma = \lambda^{-1} \gamma {\rm e}^\phi \nn \\
{\rm e}^\alpha {\rm e}^\phi & = & {\rm e}^\phi {\rm e}^\alpha \qquad
\beta \gamma = \gamma \beta\,.
\label{HW}
\eea

In the limit $p = q$, or $Q = q$ and $\lambda = 1$, ${\cal D}$ becomes a
central element of ${\cal A}$ and $GL_{p,q}(2)$ $\longrightarrow$
$GL_q(2)$; further, choosing ${\cal D}$ $=$ $1$ (or $\xi$ $=$ $1$) leads
to the quantum group $SL_q(2)$. In these cases, i.e., for $GL_q(2)$ and
$SL_q(2)$, with $\lambda$ $=$ $1$, $z$ drops out of the picture and hence
we may simply denote the $(2j+1)$-dimensional representation matrix as
${\cal T}^{(j)}$ which is obtained by taking $\lambda$ $=$ $1$ in
${\cal T}^{(j,\frac{1}{2})}$.  Then, for $SL_q(2)$ the matrix elements
$\{ {\cal T}^{(1)}_{mk} \mid m,k = 1,0,-1 \}$ of
${\cal T}^{(1)}$ in (\ref{Tau1abcd}), with $\lambda$ $=$ $1$ and
$\xi$ $=$ $1$, are seen to coincide with the spin-$1$ representation
functions $\{ d^1_{mk} \mid m,k=1,0,-1 \}$ given by
Nomura~\cite{N} (Note: our $q$ is Nomura's $q^{-\frac{1}{2}}$).

For $SU_q(2)$ one has to take into account further relations among
$\{ a,b,c,d \}$ due to the requirement of the existence of an involutional
antihomomorphic $*$-operation satisfying
%
\beq
T^* = \left(
\ba{cc}
a^* & c^* \\
b^* & d^*
\ea \right) = T^{-1} = \left(
\ba{cc}
d & -q^{-1}b \\
-qc & a
\ea \right)
\label{su*}
\eeq
so that
%
\beq
T = \left(
\ba{cc}
a & b \\
c & d
\ea \right) = \left(
\ba{cc}
a & -qc^* \\
c & a^*
\ea \right)
\label{suT}
\eeq
with $q \in \R \backslash \{0\}$ necessarily.  It may be noted that
the unitarity condition (\ref{su*}), or $T^*T$ $=$ $TT^*$ $=$ $\one$,
requires the relations
%
\beq
aa^* + q^2 c^*c = a^*a + c^*c = 1
\label{aa*}
\eeq
to be satisfied, besides the commutation relations (\ref{comrel}).

Let us now generalize the above result.  For $\Gamma^{(j,\frac{1}{2})}$,
the matrix elements are
%
\bea
\hat{J}_{\pm ,mk} & = &
Q^{\mp m+\frac{1}{2}}
\{ [j \pm m][j+1 \mp m] \}^{\frac{1}{2}}
\delta_{m,k \pm 1} \nn \\
\hat{J}_{0,mk} & = & m\delta_{mk} \qquad
\hat{Z}_{mk} = \frac{1}{2}\delta_{mk} \nn \\
   &   &  \qquad  m,k = j,j-1,\ldots ,-(j-1),-j\,.
\label{jmk}
\eea
Substituting this representation (\ref{jmk}) in (\ref{calT}) it is found,
after considerable algebraic manipulations, that one can write
%
\bea
{\cal T}^{\left( j,\frac{1}{2} \right)}_{mk} & = & \xi^{2j-1}
Q^{-\frac{1}{2}(m-k)(2j-m+k)}\lambda^{-\frac{1}{2}(m-k)(2j-m-k-1)} \nn \\
   &    &\,\times\,{\{[j+m]![j-m]![j+k]![j-k]!\}}^{\frac{1}{2}}
          \nn \\
   &    &\,\times\,\sum_{s}\,Q^{-s(2j-m+k-s)}\lambda^{-s(m-k+s)} \nn \\
   &    & \quad \quad \frac{a^{j+k-s}}{[j+k-s]!} \frac{b^{m-k+s}}{[m-k+s]!}
                \frac{c^s}{[s]!} \frac{d^{j-m-s}}{[j-m-s]!}\,.
\label{Taumk}
\eea
where $s$ runs from $\max (0,k-m)$ to $\min (j+k,j-m)$.
In the limit $\lambda$ $=$ $1$, $Q$ $=$ $q$ and $\xi$
$=$ $1$, corresponding to $SL_q(2)$ (or $SU_q(2)$), the above expression
(\ref{Taumk}) for ${\cal T}^{(j)}_{mk}$ coincides with Nomura's
expression~\cite{N} for the quantum $d$-function $d^j_{mk}$ (with our $q$
replaced by Nomura's $q^{-\frac{1}{2}}$ as already noted).  Nomura~\cite{N}
has also noted the RTT-relation (\ref{RcalTT}) for the representation
matrix ${\cal T}^{(j)}$ of $SU_q(2)$.  When $\xi$ is not taken specifically
to be unity the above matrices
$\{ {\cal T}^{(j)} \mid j=0,\frac{1}{2},1,\frac{3}{2},\ldots \}$
(\ref{Taumk}), with $Q$ $=$ $q$ and $\lambda$ $=$ $1$, provide the
representations of $GL_q(2)$ (see~\cite{F} which gives these
representations with an equivalent expression for the rhs of (\ref{Taumk})).
The rhs of (\ref{Taumk}) can be rearranged in several ways and so there
exist various equivalent expressions for ${\cal T}^{(j)}_{mk}$ in terms of
different $q$-special functions for the cases of $SL_q(2)$ and $SU_q(2)$
(see~\cite{VS,K,Ma-et,KK,N}).

When $z \neq \frac{1}{2}$, the
$\Gamma^{(j,z)}$-representation of $U_{p,q}(gl(2))$ is given by
\bea
\hat{J}_{\pm ,mk} & = & \lambda^{z-\frac{1}{2}}Q^{\mp m+\frac{1}{2}}
\{ [j \pm m][j+1 \mp m]\}^{\frac{1}{2}} \delta_{m,k\pm 1} \nn \\
\hat{J}_{0,mk} & = & m\delta_{mk} \qquad
\hat{Z}_{mk} = z\delta_{mk} \nn \\
  &   &  \quad m,k = j,j-1,\ldots ,-(j-1),-j\,.
\label{4.10z}
\eea
Substituting this representation~(\ref{4.10z}) in (\ref{calT}) one gets
\beq
{\cal T}^{(j,z)}_{mk} = \lambda^{(m-k)(z-\frac{1}{2})} \xi^{1-2z}
{\cal T}^{(j,\frac{1}{2})}_{mk} \qquad m,k = j,j-1,\ldots ,-(j-1),-j\,.
\label{4.11z}
\eeq
As is seen from~(\ref{4.11z}), for generic $z$, the one and two
dimensional representations are respectively given by
\beq
{\cal T}^{(0,z)} = \xi^{-2z} = {\cal D}^{z}\,,
\label{1drep-z}
\eeq
and
\beq
{\cal T}^{(\frac{1}{2},z)} = {\cal D}^{z-\frac{1}{2}}\left(
\ba{cc}
a & \lambda^{z-\frac{1}{2}}b \\
\lambda^{-(z-\frac{1}{2})}c & d
\ea \right)\,.
\label{2drep-z}
\eeq
As already noted at the end of Section 3, it is seen that
${\cal T}^{(\frac{1}{2},z)}$ corresponds to the fundamental
$T$-matrix of $GL_{p',q'}(2)$ with $p'$ $=$
$p^{z+\frac{1}{2}}/q^{z-\frac{1}{2}}$ and $q'$ $=$
$q^{z+\frac{1}{2}}/p^{z-\frac{1}{2}}$.

\section{$U_{\bar{q},q}(2)$}
\setcounter{equation}{0}

\noindent
It is not possible to have $SU_{p,q}(2)$ with $p \neq q$ and for $SU_q(2)$
it is necessary that $q$ $\in$ $\R \backslash \{0\}$.  But, we can
have $U_{p,q}(2)$ with $p$ $=$ $\bar{q}$ (complex conjugate of $q$);
we can have $U_{\bar{q},q}(2)$ for any $q$ $\in$
$\C \backslash \R $.  For $q$ $\in$
$\R \backslash \{0\}$ one gets $U_q(2)$ of which $SU_q(2)$ is the
special case corresponding to unit value for the quantum determinant.  To
see the features of $U_{\bar{q},q}(2)$ and $U_q(2)$ one has to study the
consequences of imposing the unitarity condition on $GL_{p,q}(2)$
(see~\cite{CJ2} for some useful details in this regard).

The fundamental $T$-matrix of $U_{\bar{q},q}(2)$, for $q$ $=$
$|q|{\rm e}^{{\rm i}\theta}$, is given by
%
\beq
T = \left(
\ba{cc}
a & b \\
c & d
\ea \right) = \left(
\ba{cc}
a & -\bar{q}\,{\cal D}c^* \\
c & {\cal D}a^*
\ea \right) = \left(
\ba{cc}
a & -qc^*{\cal D} \\
c & a^*{\cal D}
\ea \right)
\label{Tqq}
\eeq
with
%
\bea
ac & = & \bar{q}ca \quad a{\cal D} = {\cal D}a \quad
ac^* = qc^*a \quad {\cal D}c^* = {\rm e}^{2{\rm i}\theta}c^*{\cal D}
\quad cc^* = c^*c \nn \\
{\cal D}^*{\cal D} & = & {\cal D}{\cal D}^* = 1 \quad
aa^* + |q|^2c^*c = 1 \quad a^*a + c^*c = 1
\label{Tqqrel}
\eea
and their $*$-conjugate relations, satisfied ($q^* = \bar{q}$).
Note that
%
\beq
TT^* = T^*T = \one
\label{TT*}
\eeq
in view of the relations (\ref{Tqqrel}), as required and
${\cal D}$ may be representated as ${\rm e}^{{\rm i}\varphi}$ with
$\varphi^*$ $=$ $\varphi$.

If $q \in \R \backslash \{0\}$ the above equations
((\ref{Tqq})-(\ref{TT*})) hold with $\theta$ $=$ $0$ and ${\cal D}$ is a
central element with values on the unit circle in
$\C$: one gets $U_q(2)$.  If the value of ${\cal D}$
is fixed to be unity $U_q(2)$ $\longrightarrow$ $SU_q(2)$.  It is obvious
that the representation matrices of $U_{\bar{q},q}(2)$ and $U_q(2)$ are
given by the formula (\ref{Taumk}) with the relations (\ref{Tqq}) between
$\{ a,b,c,d \}$ taken into account.

\section{Clebsch-Gordan coefficients for $U_{p,q}(gl(2))$ and
$GL_{p,q}(2)$}
\setcounter{equation}{0}

\noindent
Let ${\cal C}$ and ${\cal C}'$ be the Clebsch-Gordan matrices (CGMs) such
that ${\cal C}^{-1}\Delta {\cal C}$ and ${\cal C}'^{-1}\Delta '{\cal C}'$
are reduced representations corresponding to the coproduct $\Delta$ in
(\ref{copro}) and the opposite coproduct.  From the relation
(\ref{univR}) it is clear that ${\cal C}'$ $=$ $R{\cal C}$ where $R$ is
the $R$-matrix obtained from the universal ${\cal R}$ by substituting the
corresponding irreducible representations involved in the coproduct
(see~\cite{CGZ} for a detailed discussion on the relation between
${\cal C}$s and $R$).  Now, let us make the  following observation on the
CGMs for $U_{p,q}(gl(2))$ (or $U_{Q,\lambda}(gl(2))$): from (\ref{FdelF})
it is easy to see that
%
\beq
{\cal C}_{Q,\lambda} = F{\cal C}_{Q,\lambda =1} \quad \quad
{\cal C}'_{Q,\lambda} = F^{-1}{\cal C}'_{Q,\lambda =1}\,.
\label{CFC}
\eeq
Explicitly writing, one has the expressions for the $(p,q)$ (or
$(Q,\lambda)$) CGCs as
\bea
\langle j_1 z_1 m_1 , j_2 z_2 m_2 | j z m \rangle &  =  &
\lambda^{m_1z_2-z_1m_2} \delta_{z,z_1+z_2}
\langle j_1 m_1 , j_2 m_2 | j m \rangle _Q \nn \\
\langle j_1 z_1 m_1 , j_2 z_2 m_2 | j z m \rangle ' &  =  &
\lambda^{z_1m_2-m_1z_2} \delta_{z,z_1+z_2}
\langle j_1 m_1 , j_2 m_2 | j m \rangle '_Q
\label{explicitCGC}
\eea
where $\{ \langle j_1 m_1 , j_2 m_2 | j m \rangle _Q \}$ and
$\{ \langle j_1 m_1 , j_2 m_2 | j m \rangle '_Q \}$ are the
$Q$-CGCs of $U_Q(gl(2))$ corresponding respectively to the
coproducts $\Delta_{Q,\lambda =1}$ and $\Delta '_{Q,\lambda =1}$
(see (\ref{coalg})).

It is particularly interesting to consider the coproducts for
$U_{\bar{q},q}(u(2))$ with
$q$ $\in$ $\C \backslash \R$.  In this case,
$Q$ $=$ $|q|$ and $\lambda$ $=$ ${\rm e}^{-{\rm i}\theta}$. Hence, the
coproduct (\ref{coalg}), with Hermitian $Z$ and $J_0$, preserves also the
Hermiticity property of the pair $(J_+,J_-)$ i.e.,
$J_\pm^\dagger$ $=$ $J_\mp$.  In the physical context this implies an
addition of $q$-angular momenta of two particles, (1) and (2),
according to the rule
%
\bea
\Delta (J_\pm ) & = &
J_\pm (1)|q|^{-J_0(2)}{\rm e}^{\mp {\rm i}\theta Z(2)} +
|q|^{J_0(1)}{\rm e}^{\pm {\rm i}\theta Z(1)}J_\pm (2) \nn \\
\Delta (J_0) & = & J_0(1) + J_0(2) \qquad
\Delta (Z) = Z(1) + Z(2)
\label{NEWPHASE}
\eea
with some `phases' which may somehow be irremovable.  In fact, one can
even have $|q|$ $=$ $1$ i.e., a modified addition rule for ordinary angular
momenta with a new additive `phase' quantum number.  This aspect of the
quantum algebra $U_{\bar{q},q}(u(2))$ may be worth probing further,
particularly, in view of the interest in physical applications of
$U_{p,q}(u(2))$ (see~\cite{BMK}).

Let us now look at the direct product representations of the quantum
group $GL_{p,q}(2)$.  From (\ref{calTT}) it follows that
%
\bea
{\cal T}_1{\cal T}_2 & = &
{{\cal E}xp}_{Q^{-2}}\{ \gamma \Delta (\hat{J}_-)\}
{\rm exp}\{ \alpha (\Delta (\hat{J}_0) +
\Delta (\hat{Z})) \} \nn \\
  &   &  \ \ \ \ \times\,{{\cal E}xp}_{Q^2}
\{ \beta \Delta (\hat{J}_+) \}
\nn \\
{\cal T}_2{\cal T}_1 & = &
{{\cal E}xp}_{Q^{-2}}\{ \gamma \Delta '(\hat{J}_-)\}
{\rm exp}\{ \alpha (\Delta '(\hat{J}_0) +
\Delta '(\hat{Z})) \} \nn \\
  &   &  \ \ \ \ \times\,{{\cal E}xp}_{Q^2}
\{ \beta \Delta '(\hat{J}_+)\}
\label{TauTau}
\eea
where $\Delta$ is given in (\ref{newcoalg}) and the representations
involved in the coproducts are the relevant representations
$\{ \Gamma^{\left( j,\frac{1}{2} \right)} \}$.  It is particularly
easy to verify these relations in the case when
$\Delta$ corresponds to the direct product of two identical
$\Gamma^{\left( \frac{1}{2},\frac{1}{2} \right)}$-representations.
It is obvious that ${\cal C}^{-1}{\cal T}_1{\cal T}_2{\cal C}$ and
${\cal C}'^{-1}{\cal T}_2{\cal T}_1{\cal C}'$ will be in reduced forms.
Since, for $U_{p,q}(gl(2))$, $\Delta^{(j_1,z_1) \otimes (j_2,z_2)}
\longrightarrow \sum_{j'=|j_1-j_2|}^{j_1+j_2}\,\oplus
\Gamma^{(j',z_1+z_2)}$, the direct product representation
${\cal T}_1{\cal T}_2$ $=$ $({\cal T}^{(j_1,z_1)} \otimes
I^{(j_2,z_2)})(I^{(j_1,z_1)} \otimes {\cal T}^{(j_2,z_2)})$ (or
${\cal T}_2{\cal T}_1$ corresponding to the opposite coproduct),
where $I^{(j,z)}$ is the $(2j+1)$-dimensional identity matrix
corresponding to the unity in the $(j,z)$-representation, will be
reduced to the direct sum of the representation matrices
 $\{ {\cal T}^{(j,z_1+z_2)} | j = |j_1-j_2|, ... ,j_1+j_2 \}$ and
the corresponding CGCs are given by (\ref{explicitCGC}).

As an interesting example, let us consider the direct products of the
representations labeled by $(0,z-\frac{1}{2})$ and $(j,\frac{1}{2})$
for both $U_{p,q}(gl(2))$ and $GL_{p,q}(2)$; note that, in general,
the $(j,z)$-representation is defined by~(\ref{4.10z}) (or (\ref{newJ})).
Using the coproduct (\ref{newcoalg}) and its opposite the corresponding direct
product representations of $U_{p,q}(gl(2))$ are given by, with
$(j_1,z_1)$ $=$ $(0,z-\frac{1}{2})$ and $(j_2,z_2)$ $=$
$(j,\frac{1}{2})$,
\bea
\Delta (\hat{J}_+)_{mk} & = & Q^{-m+\frac{1}{2}}\{ [j+m][j+1-m] \}^
{\frac{1}{2}}\delta_{m,k+1} \nn \\
\Delta (\hat{J}_-)_{mk} & = & \lambda^{2z-1} Q^{m+\frac{1}{2}}
\{ [j-m][j+1+m] \}^{\frac{1}{2}}\delta_{m,k-1} \nn \\
\Delta (\hat{J}_0)_{mk} &= & m \delta_{mk} \qquad
\Delta(\hat{Z})_{mk} = z \delta_{mk} \nn \\
  &    &  \quad m,k = j,j-1, \ldots ,-(j-1),-j
\label{Deltarep-alg}
\eea
and
\bea
\Delta '(\hat{J}_+)_{mk} & = & \lambda^{2z-1} Q^{-m+\frac{1}{2}}
\{ [j+m][j+1-m] \}^ {\frac{1}{2}}\delta_{m,k+1} \nn \\
\Delta '(\hat{J}_-)_{mk} & = & Q^{m+\frac{1}{2}}
\{ [j-m][j+1+m] \}^{\frac{1}{2}}\delta_{m,k-1} \nn \\
\Delta ' (\hat{J}_0)_{mk} &= & m \delta_{mk} \qquad
\Delta '(\hat{Z})_{mk} = z \delta_{mk} \nn \\
  &    &  \quad m,k = j,j-1, \ldots ,-(j-1),-j\,.
\label{Delta'rep-alg}
\eea
Using these representations in (\ref{TauTau}) above it is seen that the
corresponding direct product representations of
$GL_{p,q}(2)$ are given by
\beq
{\cal T}_1{\cal T}_2 = {\cal D}^{z-\frac{1}{2}}
{\cal T}^{(j,\frac{1}{2})} \qquad
{\cal T}_2{\cal T}_1 = {\cal T}^{(j,\frac{1}{2})}
{\cal D}^{z-\frac{1}{2}}\,.
\label{Deltarep-group}
\eeq
It is easy to verify that the direct product representations of
$U_{p,q}(gl(2))$ given by~(\ref{Deltarep-alg})
and~(\ref{Delta'rep-alg}) can be `reduced' to (or made equivalent
to, in this case) the defining representation~(\ref{4.10z}) using
the CGCs obtained from~(\ref{explicitCGC}) by taking the $Q$-CGMs to
be identity matrices.  The same CGMs are seen to `reduce' also the
direct product representations of $GL_{p,q}(2)$ given
by~(\ref{Deltarep-group}) to the defining
representation~(\ref{4.11z}).  It may also be noted that the direct
product representations of $GL_{p,q}(2)$ given
by~(\ref{Deltarep-group}) satisfy the $RTT$-relation (\ref{RcalTT}) with a
$(2j+1)$-dimensional `$R$-matrix' with elements $\{ R_{mk} =
\lambda^{(2z-1)m}\delta_{mk} | m,k = j,(j-1), \ldots ,-(j-1),-j \}$
in accordance with (\ref{FRF}).

\section{Conclusion}

\noindent
Before closing, we may mention a few related points.

The converse of the exponential map, namely, the passage $GL_{p,q}(2)$
$\longrightarrow$ $U_{p,q}(gl(2))$ using the
representation of $GL_{p,q}(2)$ close to identity, for
infinitesimal values of the group parameters $\{ \alpha ,\beta ,\gamma
,\delta \}$, follows by writing ${\cal T}$ $\approx$
$1 + \gamma \hat{J}_- + \alpha (\hat{J}_0 + \hat{Z}) +
\delta (\hat{J}_0 - \hat{Z}) + \beta \hat{J}_+$.  Thus, if
one can obtain the representations of $GL_{p,q}(2)$ by some method
directly, then, the representations of its infinitesimal generators
$\{ \hat{J}_0,\hat{J}_\pm ,\hat{Z} \}$ forming the dual
algebra $U_{p,q}(gl(2))$ can be derived.  This is how Finkelstein~\cite{F}
obtains the relationship $GL_q(2)$ $\longrightarrow$ $U_q(gl(2))$,
independent of~\cite{FG}, using theory of invariants to derive an
equivalent form of the representation matrix (\ref{Taumk}) for the case
$p=q$.

Finkelstein's analysis of the representations of $GL_q(2)$~\cite{F}
is motivated by the construction of a $GL_q(2)$ Yang-Mills theory in which
one would regard the nonabelian group parameters
$\{\alpha (x),\beta (x),\gamma (x),\delta (x) \}$ (the coproduct rules of
which specify the group multiplication law as pointed out in~\cite{MV}) as
space-time fields.  Akulov {\em et al.}~\cite{AGG}
have considered the differential calculus of the group parameter space for
$SL_q(2)$ and studied a related field theory model.  The problem of
realization of the group parameters as dependent on continuous
classical variables (like space-time) has been addressed recently by
Volovich~\cite{V} at the level of the variables $\{ a,b,c,d \}$ (see
also~\cite{CJ2}).
Let us observe an example of such a realization based on the
relations (\ref{a=}) and (\ref{Lie}):  Using the well known Bogoliubov
transformation of a pair of commuting sets of boson operators, we
can write
\begin{eqnarray*}
&&\alpha =(\rho - \theta )\psi_1^\dagger (x)\psi_1(x)
+(\rho + \theta )\psi_2^\dagger (x)\psi_2(x),\\
&&\beta=  \psi_1^\dagger (x),\qquad \gamma=\psi_2^\dagger (x), \\
&&\delta =(\rho + \theta )\psi_1^\dagger (x)\psi_1(x)+(\rho -
\theta )\psi_2^\dagger (x)\psi_2(x),
\end{eqnarray*}
with $\psi_1(x)$ $=$
$({\rm cosh}x)a_1$ $-$ $({\rm sinh}x)a_2^\dagger$, $\psi_2(x)$ $=$
$({\rm cosh}x)a_2$ $-$ $({\rm sinh}x)a_1^\dagger$, $x$ $\in$ $\R$,
$[a_1,a_1^\dagger ]$ $=$ $1$, $[a_2,a_2^\dagger ]$ $=$ $1$,
$[a_1,a_2]$ $=$ $0$, $[a_1,a_2^\dagger ]$ $=$ $0$.
In the context of building gauge theories based on
quantum groups related to $GL(2)$ it is interesting to observe that in the
case of $GL_{p,q}(2)$ the two-dimensional vector spaces carrying the
fundamental representation ($T$) have commuting components if $p$ $=$ $1$
just like the Hilbert space of a two-level ordinary quantum mechanical
system.  Besides the interpretation of
the matrix elements of the representations of $SU_q(2)$ as wavefunctions of
quantum symmetric tops~\cite{N}, generalization of quantum dynamics
based on the representations of $SU_q(2)$ have also been
considered~\cite{A-et}.  Since having one more parameter would provide
more flexibility in model building, we believe that the study of
representations of $GL_{p,q}(2)$ should prove useful.  Recently,
generalization of the exponential map for the quantum supergroup
$GL_{p,q}(1|1)$ has also been obtained~\cite{CJ3}.

There are several approaches to quantum groups (or quantum matrix
pseudogroups) and quantum algebras (or quantized universal enveloping
algebras) which are in duality with the quantum groups in the Hopf
algebraic sense (see, e.g.,~\cite{M,DHL} for reviews of the subject).
In the case of the quantum group
$GL_{p,q}(2)$~\cite{Ku,De-et,S1,T} the structure of its dual Hopf algebra
$U_{p,q}(gl(2))$ is known from the studies by
Schirrmacher et al.~\cite{SWZ}, using the noncommutative differential
calculus approach, and by Dobrev~\cite{D1,D2,D3} using the approach of
Sudbery~\cite{S2}.  The recent analysis~\cite{FG} of the duality
relations between Lie bialgebras, with particular reference to
$A(GL_{p,q}(2))$ and $U_{p,q}(gl(2))$, has led to a generalization of the
well-known exponential relationship between a classical Lie group and its
Lie algebra and the explicit form of such an exponential map
has been obtained between the quantum algebra $U_{p,q}(gl(2))$ and the
corresponding quantum group $GL_{p,q}(2)$.  This relationship is given
abstractly in terms of a universal ${\cal T}$-matrix, involving both the
group parameters of $GL_{p,q}(2)$ and the generators of $U_{p,q}(gl(2))$,
and for particular representations of $U_{p,q}(gl(2))$  this universal
${\cal T}$-matrix gives the representations of $GL_{p,q}(2)$.  Using this
Fronsdal-Galindo formalism we have derived explicitly the
finite-dimensional representations of $GL_{p,q}(2)$, by exponentiating
directly the well-known
$(2j+1)$-dimensional irreducible representations of $U_{p,q}(gl(2))$, and
the earlier results on the representations of $GL_q(2)$, $SL_q(2)$ and
$SU_q(2)$ are found to be special cases in the appropriate limits.  We
have also derived the CGCs for the quantum algebra $U_{p,q}(gl(2))$
and noted that the same CGCs can be used for the Clebsch-Gordan
reduction of the direct product representations
of the quantum group $GL_{p,q}(2)$.

\section*{Acknowledgements}

\noindent
We have pleasure in thanking Prof. T.D. Palev for stimulating discussions.
One of us (RJ) is thankful to Prof. Guido Vanden Berghe for the kind
hospitality at the Department of Applied Mathematics and Computer Science,
University of Ghent.  RJ wishes to thank Prof. K. Srinivasa Rao for
helping him in many ways during his stay abroad.  He also acknowledges
the benefit of discussions with Prof. K. Srinivasa Rao, Dr. V. Rajeswari
and Dr. R. Chakrabarti, some time ago, on the problem of CGCs for
$U_{p,q}(gl(2))$.  This research was partly supported by the E.E.C
(Contract No. CI1*-CT92-0101).

\end{document}